\documentclass[journal=mamobx,manuscript=article]{achemso}
\usepackage{amsmath}
\usepackage{mciteplus}
\usepackage{subfig}
\usepackage{graphicx}

\usepackage{xcolor}
\author{Syed Shuja Hasan Zaidi}
\affiliation{Department of Physics, Indian Institute of Technology Jodhpur, Karwar 342030, India}
\author{Madhu Priya}
\affiliation{Department of Physics, Birla Institute of Technology Mesra, Ranchi 835215, India}
\author{Sanjay Puri}
\affiliation{School of Physical Sciences, Jawaharlal Nehru University, New Delhi 110067, India}
\author{Prabhat K. Jaiswal}
\affiliation{Department of Physics, Indian Institute of Technology Jodhpur, Karwar 342030, India}
\email{prabhat.jaiswal@iitj.ac.in}

\title{Surface-directed spinodal decomposition in binary fluid mixtures on an amorphous wall: A molecular dynamics study}

\abbreviations{IR,NMR,UV}

\begin{document}


\begin{tocentry}
    \includegraphics[width=0.95\linewidth]{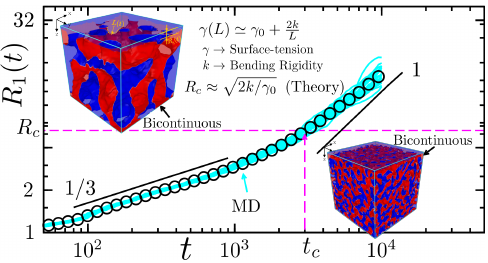}
\end{tocentry}

\newpage
\begin{abstract}
	We present molecular dynamics (MD) results to discuss wetting kinetics in binary fluid mixtures ($A:B=50:50$) undergoing surface-directed spinodal decomposition (SDSD) on an amorphous wall. Our simulations show the formation of a wetting layer rich in the preferred $A$-type particles and bicontinuous domain morphology in the bulk. In addition, the mixture maintains connectivity between the bulk and the wetting layer through $A$-rich tubes throughout the depletion region. The wetting layer thickness coarsens as a power law, $R_1(t)\sim t^{\alpha}$, with two distinct growth regimes of $\alpha=1/3$ and $\alpha=1$ active for at least a decade. The computed crossover time for $\alpha=1/3 \to 1$ equaled the reported bulk crossover time, and the corresponding crossover length scale $R_c$ agrees well with the expression $\Lambda = \sqrt{2k/\gamma_0}$ given by Scholten et al.~[\emph{Macromolecules}2005, 38, 3515] for bicontinuous domains in aqueous polymer mixtures in the presence of only one dominant length scale. This agreement supports a hydrodynamic picture of diffusive growth for the interconnected wetting layer and bulk domains, where the bending contribution ($k$) of curvature-dependent $AB$ interfacial tension ($\gamma$) governs small-scale coarsening, producing $t^{1/3}$ growth. For length scales beyond $\Lambda$, capillary flows yield the viscous hydrodynamic regime ($\sim t$). Our results show no orientational effects on the domain coarsening parallel and perpendicular to the wall, contrasting many continuum models, including combinations with Flory-Huggins theory. 
\end{abstract}

\section{Introduction} Phase separation quenched deep into the miscibility gap proceeds via spinodal decomposition (SD), where all concentration fluctuations lower the mixture free energy.\cite{VS09,tanaka15naturec,bray02,onuki02,JJ20macro} SD has an early stage in which inhomogeneities develop into interfaces and a late-stage domain-coarsening regime in which the total interfacial area decreases as the characteristic domain size $L(t)$ grows. 
When a single length scale $L$ dominates, the domain morphology shows dynamical scaling and $L\sim t^\alpha$, with the growth exponent $\alpha$ identifying the transport mechanism. The volume fraction $v$ of the minority phase leads to distinct morphologies and exponents:\cite{bray02,huse86prb,siggia79,KD74,binder77prb,IV61} evaporation-condensation (EC) or Ostwald ripening for $v < 1\%$ ($\alpha=1/3$, Lifshitz-Slyzov\cite{IV61}), droplet coalescence or Brownian coagulation (BC) for $v > 1\%$ ($\alpha=1/3$, Binder and Stauffer\cite{binder77prb,KD74,siggia79}), and hydrodynamic flow (HF) due to capillary pressure for $v\approx 50\%$ ($\alpha=1$, Siggia\cite{siggia79} for bicontinuous morphology). Interestingly, fixed $v$ mixtures (that is, a domain geometry) also exhibit crossovers from one dominant regime to the other; the crossover from $\alpha=1/3$ to $\alpha=1$, and vice versa, is common in the literature for off-critical ($v\ne 50\%$) and critical ($v\approx 50\%$) mixtures.\cite{VS09,siggia79,bray02,tanaka15naturec,KSpre23,ERD11jpcb,Bhat06jpcm,NASA,NAA01macro34,ELE05macro38,GDVARH03biomacro,NJJKH96physica} For example, the bicontinuous domain coarsening in an aqueous polymer mixture (gelatin/maltodextrin) shows a $1/3 \to 1$ crossover driven by HF, with a crossover length scale $L_c = \Lambda = \sqrt{2k/\gamma_0}$ obtained by balancing the bending rigidity $k$ and the surface tension $\gamma_0$ of the domain interfaces.\cite{NAA01macro34,ELE05macro38} This crossover accounts for the bending energy contributing to the interfacial energy, which diverges for $L<\Lambda$ as the surface tension becomes curvature-dependent, $\gamma(L)\propto 1/\gamma^2$. This theme of exponents and crossovers is common to fluid mixtures such as colloid-polymers,\cite{ERD11jpcm23,Bhat06jpcm,NASA,NJJKH96physica} polymer blends,\cite{NAA01macro34,tanaka15naturec} and molecular liquids.\cite{tanaka15naturec,tanaka01} However, a high bending rigidity $(k \sim 10^2k_BT)$ and low surface tension for colloid and polymer structures yield micron-scale $L_c$, 
readily visible by optical microscopy.\cite{NAA01macro34,NASA,ELE05macro38,Bhat06jpcm,ERD11jpcm23}

In practice, phase separation usually occurs in the presence of boundaries rather than in a boundaryless (bulk) setting, so wall-induced effects can alter the overall behavior of thin films\cite{SSJK06prl,SSJK06pre,RPS13jcp,tanaka00jpcm,SMJ97,MHG00,MG03} and the local behavior in semi-infinite setups.\cite{puri05,tanaka01,MM10,SSSSR17epl,SSPSP24jcp,ZJPP22pre,GAJ94pre,GEF94,HA97pre} These wall effects extend well beyond the molecular length scales \cite{RLE91,CWA95pre,tanaka01} and arise whenever a substrate has affinity for one component (say $A$), producing surface wetting by that phase. In the case of complete wetting (CW), the $B$ phase is expelled from the wall and a macroscopically thick $A$-rich wetting layer forms; this generates oscillatory composition waves normal to the wall that decay deep into the bulk. These composition waves are the SDSD waves, consisting of the wetting layer of the preferred phase on the wall, followed by the depletion layer of the nonpreferred phase. This wetting-spinodal demixing interplay is therefore termed surface-directed spinodal decomposition (SDSD).\cite{puri05,KSS10,tanaka01} 
The SDSD waves break the phase separation symmetry perpendicular to the wall, resulting in different dynamical behavior of length scales, $L_{||}(z,t)$ and $L_{\perp}(y,t)$~or~$L_{\perp}(x,t)$, proprtional to the characteristic size of domains parallel and perpendicular to the wall. In addition, the thickness of the wetting layer [$R_1(t)$], given by the first zero crossing of SDSD waves, becomes an experimentally accessible length scale and has gone through intense scrutiny in experiments,\cite{MRA95,RLE91,MG03,krausch95, PA91,APF92,BCA93,Michels11,YDL11} theory,\cite{SK92,troian93prl,troian92,HPW97,RR90,KH91,marko93} and simulations.\cite{SY88,SH97,puri05,binder83,GA92,SK01,SK94,PSS12,SSJ01,SSJ05,SSJ06-pre,bray02,APF21,toxvaerd99prl,HA97pre,PWA94prl,HT00,tanaka01,SAR20,PPS20,PPS20-2,MPJ16} 

Jones et al.\cite{RLE91} developed a depth profiling technique to observe SDSD waves in polymer blends, which became a stepping stone for later exploration of SDSD. Later experimental studies on fluid mixtures have reported the growth of the wetting-layer thickness as $R_1(t)\sim t^{\alpha}$, with growth exponents identifying with the similar transport mechanisms as seen for bulk mixtures, for given domain geometries.\cite{puri05,tanaka01} The reported $\alpha$ values are close to $1/3$ for blends of higher molecular weight polymers.\cite{puri05,RLE91,MG03,FR92,GCE93,GEF94,MRA95,MTJR97macro,JE13macro,MTJR97macro} However, for fluid mixtures such as oligomeric polymer solutions,\cite{PA91,APF92,FP89} molecular liquids,\cite{NC81,YW79,PRFD87pra,tanaka93-2,tanaka93-3,tanaka01} and colloidal-polymer mixtures,\cite{ERD11jpcb} HF dominates over diffusive mechanisms; thus, $\alpha\approx 1$ is reported frequently for critical mixtures when the wetting material flows from the bulk to the wetting layer through fluid tubes. Furthermore, bulk crossovers are also observed for $R_1(t)$ in mixtures with interconnected morphology, especially $1/3\to 1$ ($1\to 1/3$ when the fluid tubes disappear, disconnecting the wetting layer from the bulk) in molecular liquids\cite{NC81,YW79,PRFD87pra,tanaka93-2,tanaka93-3,tanaka01} and aqueous colloidal suspensions.\cite{ERD11jpcb} However, an important observation was made in colloid-polymer mixtures\cite{ERD11jpcb} where the crossover of $1/3\to 1$ in the wetting-layer growth occurs later than the bulk crossover. Apart from the usual bulk crossovers, SDSD in molecular liquids\cite{NC81,YW79,CWA95pre,APF92} shows a unique crossover to a fast mode with $\alpha=3/2$ at an early stage, a value greater than even the $\alpha=1$ of HF in the late regime. The possible mechanism behind such explosive growth has been debated extensively,\cite{troian92,troian93prl,troian94,PWA94prl,tanaka01} with Tanaka's\cite{tanaka01} explanation based on HF under strong wetting conditions in which the preferred phase spreads on the wetting layer in a pseudo-two-dimensional way. Furthermore, colloid-polymer mixtures\cite{ERD11jpcb} have also qualitatively shown that the domains of the preferred phase are elongated parallel to the wall, due to increased friction along the wall. 

Puri and Binder (PB)\cite{puri05,SKH97} gave the first successful diffusion-limited model (model B in the Hohenberg and Halperin universality classes\cite{PB77}) of SDSD by extending the Cahn-Hilliard theory of SD in binary alloys to mixtures under long-range potentials ($V \sim z^{-n}$, $n$ being the potential range) from semi-infinite surfaces. In the coarsening regime, PB found that the wetting-layer growth is initially driven by the surface potential as $R_1(t)\sim t^{1/(n+2)}$, which in the late stages becomes $1/3$ governed by the negative chemical gradient from the flat wetting layer and the bulk domains. Later, model H by Tanaka in 3D\cite{tanaka01} and Chan and Chakrabarty in 2D\cite{HA97pre} gave HF insight into $R_1(t)$ growth with $t^{1/3} \to t^1$ and $t^{1/3} \to t^{1/2}$, due to 2D capillarity, crossovers in the late stages; however, Tanaka mainly focused on confinement effects,\cite{tanaka01,tanaka93,tanaka93-2,tanaka93-3,HT00,HT95} which totally alter phase behavior due to multiple SDSD waves from different wall faces. Model B was extended to polymer systems by combining it with the well-known Flory-Huggins theory in 2D systems.\cite{LX06macro} Their results show dynamical scaling of SDSD waves and gave three detailed $R_1(t)$ coarsening regimes determined by the quench-depth, with $R_1(t)\sim t^{1/3}$ (LS growth law) for the deepest quench. The 2D results of model H and the polymeric numerical analysis demonstrated that the evolution of the domain morphology in parallel cross sections in the wall vicinity obeys the LS growth law ($L_{||}\sim t^{1/3}$); furthermore, the orientational effects of the walls introduce the disparity between the dynamical behavior of these cross sections ($L_{||} \ne L_{\perp}$). The wetting-layer formation mechanism is also explored by Lattice-Boltzmann methods (LBM),\cite{APF21} where the results of the wetting-layer growth reported the crossover $1/3\to 1$. However, they did not account for the different dynamical behaviors of $L_{||}$ and $L_{\perp}$.

	Molecular dynamics (MD) studies aimed at investigating SDSD have reported different growth regimes and crossovers of $R_1(t)$, for various geometries and confinements, such as semi-infinite geometry,\cite{puri05,PSS12,PSS10,PSS12-epl,PWA94prl, PWA93,ZJPP22pre} thin films,\cite{puri05,SSJK06prl,SSJK06pre,RPS13jcp} etc. 
    All these MD works share a common feature of representing the surface by an integrated Lennard-Jones potential. The integrated potential smoothens the surface's roughness to a vanishing scale, unnoticed by the fluid particles. This smooth surface model exhibits stacking of particles in layers near the surface,\cite{PWA93,PWA94prl,SSJK06prl}, which induces anisotropic density fluctuations in the direction parallel and perpendicular to the wall. 
    Furthermore, some studies have used structured walls with face-centered cubic (FCC) lattice to investigate SDSD in thick films using MD,\cite{PWA93} and in thin films and nanopores via dissipative particle dynamics (DPD) simulation.\cite{MM10,SSSSR17epl} However, the crystalline state of the wall also produced similar layering effects. 

In this work, we have focused on MD simulations by eliminating the layering effects of the integrated wall, replacing it with an amorphous wall composed of particles. The latter increases the surface roughness to the fluid particle scale. Our effort of making rougher walls has been fruitful, and we could capture the wetting-layer growth for larger lengths and decade-long time scales. Relaxed fluid density near the walls means removal of the particle layering and consequently lowers the chances of wall-induced crystallization of the mixture. 
Subsequently, we investigate the wetting-layer growth, extracting growth exponents, crossovers, and discussing key questions such as the existence of \emph{fast-mode kinetics}, matching growth regimes and crossovers against experiments on fluid mixture for critical mixture compositions leading to bicontinuous domain morphology. In addition, we critically assess the anisotropic domain coarsening near the wetting wall by quantifying $L_{||}(t)$ and $L_{\perp}(t)$ in cross sections parallel and perpendicular to it. 

The universality of the coarsening regimes and crossovers between experiments and simulations, and across fluid mixtures with different interaction strengths and inherent lengths and time scales ($nm \to \mu m$ and milliseconds $\to$ seconds) states that the phase separation kinetics does not depend on the exact nature of interactions. Some modest interaction models, such as those employed in our MD simulations, could readily produce domain coarsening kinetics and crossovers on micron-length scales observed in experiments with different fluid mixtures. MD simulations with minimal input provide greater flexibility than experiments and phenomenological lattice models with imposed fields to explore the microscopic mechanisms behind the crossovers. Thus, our MD simulations bridge the gap between the experimental observations (especially polymer mixtures) and phenomenological theories to deeply understand the SDSD mechanism and the interplay between the wetting kinetics and SD. The SDSD is inevitable during the fabrication of metal oxide semiconductors\cite{XWD09prb} and organic electronics\cite{JDM15,SMJ97,Michels11} via processes like atomic layer deposition and solvent evaporation (drying). Therefore, it finds utilization in developing high-dielectric gate transistors,\cite{XWD09prb} bulk heterojunctions in organic electronics,\cite{LEV05,JRI10,KDB08,Chen19,YDL11,Michels11,HYX20,CJP17} etc, as it guides the formative domain morphology to be frozen for a productive active matter of suitable percolation length based on exciton relaxation length scales.

\section{Model and Simulation Details}
We simulated surface-directed spinodal decomposition (SDSD) on an amorphous substrate using molecular dynamics (MD) with the highly parallel LAMMPS engine\cite{plimpton95}. The simulation box has volume $L_x\times L_y\times L_z$ and is periodic in $x,y$; the fluid is confined between two walls along $z$. The fluid particles interact via the standard 12-6 Lennard-Jones (LJ) potential
\begin{equation}\label{eq3:LJ}
U_{\alpha \beta}(r) = 4\epsilon_{\alpha\beta}\left[ \left(\frac{\sigma_{\alpha \beta}}{r}\right)^{12}- \left(\frac{\sigma_{\alpha \beta}}{r}\right)^6\right],
\end{equation}
with $\epsilon_{AA}=\epsilon_{BB}=2\epsilon_{AB}=\epsilon$ and $\sigma_{AA}=\sigma_{BB}=\sigma_{AB}=\sigma$ to model a symmetric binary mixture, following previous MD studies of SD and SDSD.\cite{SSS10,SSS12,PSS10,PSS12,PSS12-epl} The bottom wall's preference for the $A$-component is fixed by setting $\epsilon_{AS}=\epsilon$ and $\sigma_{AS}=\sigma$.

Neutral (purely repulsive) wall-fluid interactions use the truncated and shifted WCA form.\cite{WCApot}
\begin{equation}\label{eq3:WCA}
\phi_{\alpha\beta}(r) = 
\begin{cases}
4\epsilon_{\alpha\beta}\left[ \left(\frac{\sigma_{\alpha\beta}}{r}\right)^{12}- \left(\frac{\sigma_{\alpha\beta}}{r}\right)^6\right]+\phi_{c},\;& r<2^{1/6}\sigma_{\alpha\beta}\\
0, & r\ge \sigma_{\alpha\beta}
\end{cases}
\end{equation} 
We set $\epsilon_{AS}=\epsilon_{BS}=\epsilon$ and $\sigma_{AS}=\sigma_{BS}=\sigma$. The bottom wall preferentially wets the $A$ component while the top substrate is neutral, producing an $A$-rich wetting layer at $z=0$.

All quantities are reported in reduced units with $\epsilon=k_BT$, $\sigma$ and $m$ set to unity; the LJ potential and forces are truncated and shifted at $R_c=2.5\,\sigma$. The cubic fluid box has a side $L=128$ and is bounded within $z=0$ and $z=128$ along the $z$ direction, while the walls occupy a slab volume of $128\times128\times2$ centered at $z=-1$ and $z=129$. We fix the number density $\rho_N= N/L^3=1$ ($N=128^3$ with $N$ being the total number of fluid particles $A+B$) to avoid liquid-gas separation; for these box dimensions, the average domain size can grow up to $32\,(=L/4)$ without finite-size artifacts. We simulate a critical mixture where $N_A=N_B=N/2$ ($N_A$ and $N_B$ being the total number of $A$ and $B$-type particles). The chosen wetting parameters correspond to a first-order wetting transition, yielding a finite equilibrium wetting-layer thickness that saturates in our finite box.

The simulation protocol is as follows. We equilibrate the fully mixed fluid at high temperature for $t<0$ with walls in place, then quench instantaneously to $T=0.7T_c$ at $t=0$ (with $T_c\simeq1.423$), and record demixing for $t>0$. Fluid motion is integrated with the velocity-Verlet algorithm while wall atoms remain fixed; temperature is controlled by a Nos\'e-Hoover thermostat to preserve hydrodynamics.\cite{MD17,allen96,SSS10,SSS12} Because phase separation is non-equilibrium, we performed 20 independent runs with different initial seeds and noise realizations.


Morphological characterization is performed by mapping the raw MD trajectory onto a cubic lattice with $L^3$ lattice points. Each lattice point holds spin up and down states of Ising-like form determined by the system's order parameter $\psi(\Vec{r},t)= (\rho_A-\rho_B)\to \pm 1$, where $\rho$ is the local density. To minimize the noise factor, we rectify $\psi(\vec{r},t)$ of each lattice point at $\vec{r}$ using the majority spin rule, where we acknowledge the contributions of $\psi$ from all its first neighbors. This method of coarse-grained mapping is similar to the numerical renormalization group technique described in Ref.\cite{SS02pre} We then compute the spatial statistics, such as structure factor and correlation functions, on the lattice as described below.\cite{SS02pre,SS10,PSS12}

\section{Results and discussion}

	\begin{figure}[!htbp]
		\centering
		\includegraphics[width=0.8\linewidth]{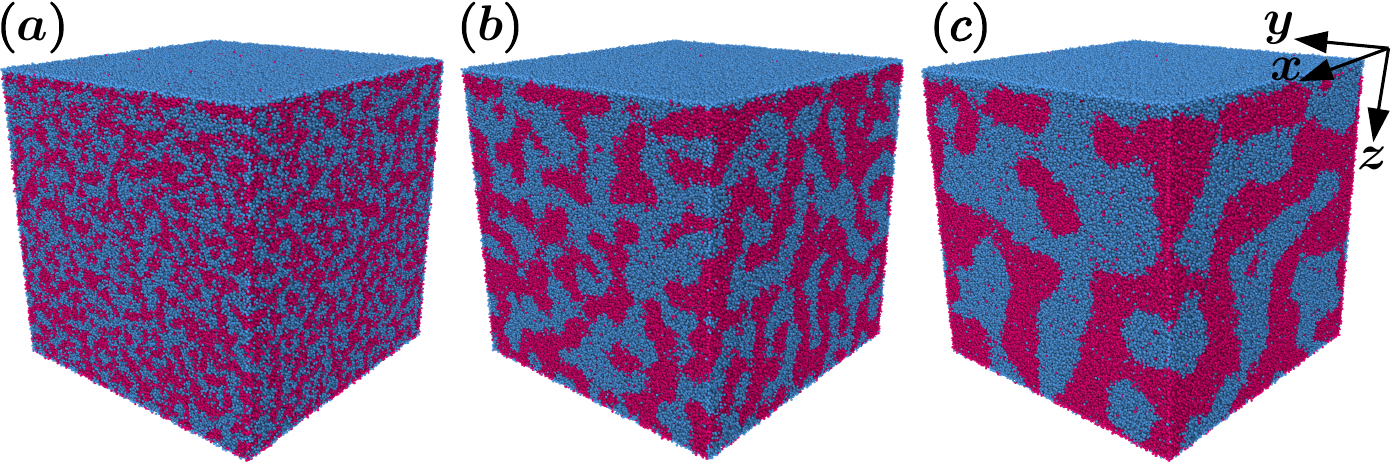}
		\caption{The time evolutionary snapshots of surface-directed spinodal decomposition in a simulation box of volume $L_x\times L_y \times L_D = 128\sigma\times 128\sigma\times 128\sigma$. Wetting occurs on the top wall located at $z=0$. The two particle types in the binary mixture ($A+B$) are shown in blue ($A$-type) and pink ($B$-type) colors. The snapshots shown here are for the time $t>0$ for a single quench to $T=0.7T_c$ at $t=0$, where $T_c(\simeq 1.423)$ is the bulk critical temperature. The snapshots belong to times $(a)\;\; t=200$, $(b)\;\; t=500$, and $(c)\;\; t=3000$.}
		\label{fig:figure1}
	\end{figure} 

\begin{figure}[!htbp]
		\centering
		\includegraphics[width=0.5\linewidth]{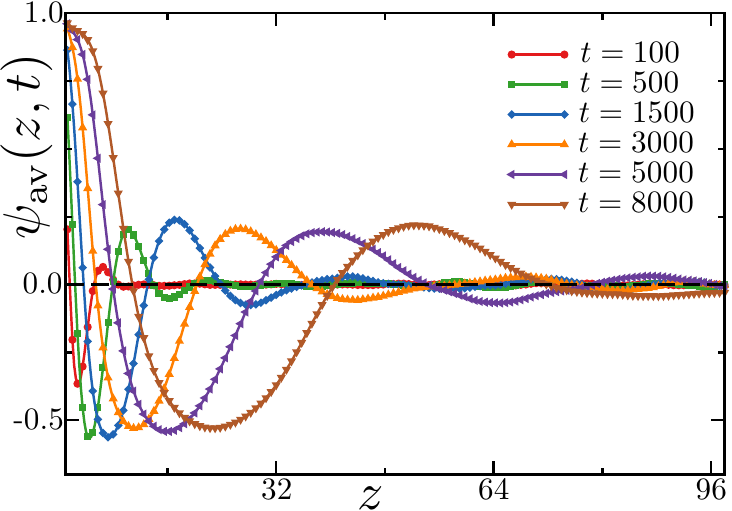}
		\caption{Laterally averaged order-parameter profiles, $\psi_{\text{av}} (z,t)$, with time specified. The profiles are constructed using a slab width of $\Delta z=0.5$. }
		\label{fig:figure2}
	\end{figure}

We begin with the illustrations of surface-directed spinodal decomposition (SDSD) presented as snapshots in Fig.~\ref{fig:figure1}. These snapshots belong to different growth regimes of $R_1(t)$ and exhibit crucial morphological features unique to a coarsening process. Starting with the top wall at $z=0$, getting coated with $A$-type particles for $t\ge 100$, the subsequent snapshots display the coarsening of the wetting layer and bulk domains together. Adjacent to the wetting layer, a region depleted in $A$-type particles is formed, called a depletion layer, as demonstrated by the laterally-averaged order parameter profiles $\psi_{\text{av}} (z,t)$ in Fig.~\ref{fig:figure2}. These profiles are computed parallel to the wall as \begin{equation}\label{eq5:psiav}
	\psi_{\text{av}} (z,t) = \frac{1}{L_x\times L_y} \int_{0}^{L_x} \int_{0}^{L_y} \psi(\;\vec{r},t\;)\; dx\;dy,
	\end{equation} 
      which is also averaged over the ensemble of different simulation runs with independent starting configurations. This quantitative analysis is much credited to the depth-profiling techniques of mixture compositions from experiments.\cite{GCE93,RLE91} The numerical adaptation of the above equation is straightforward and is done by partitioning the fluid box into cuboidal slabs of size $128\times 128\times \Delta z$, where $\Delta z=0.5$ is the slab width in the $z$-direction, positioned at depth $z$. For the depletion layers, $\psi_{\text{av}} (z,t)$ values are below the global composition value $\psi_0= (N_A-N_B)/(N_A+N_B)=0$ as total $A$-type particles are less than $B$-type particles in those slabs for all $z$ lying in the depletion region. Similarly, $\psi_{\text{av}} (z,t)$ in the wetting layers $\psi_{\text{av}} (z,t)> \psi_0$, which saturates to $+1$ near the surface, as the nonwetting components are completely expelled from the surface. Also, the late stage ($t>1500$) profiles show a neck near the wetting wall, which later develops into a shoulder, commonly observed in profiles from experiments and phase-field simulations. In particular, unlike the wetting layer, the depletion layers are not devoid of the wetting component, since $\psi_{\text{av}} (z,t)$ in them does not decay to close to $-1$ over time, and shows a minimum value corresponding to a global composition of $\psi_0=-0.5$, i.e., $A:B=25:75$. The degree of depletion of the wetting component $A$ agrees with the results obtained in numerical simulations of CHC theory combined with the Flory-Huggins-de Gennes theory for polymeric mixtures. However, this behavior is not replicated in the simulations of model-B by Puri-Binder, where $\psi_{\text{av}} (z,t)$ is strictly conserved within the bilayer of the wetting and depletion layers; thus, $\psi_{\text{av}} (z,t)$ in the depletion layer decays to $-1$ as it grows to $+1$ in the wetting layer with net composition equal to $\psi_0$ for the bilayer. 
   
  	\begin{figure}[!htb]
		\centering
		\includegraphics[width=\linewidth]{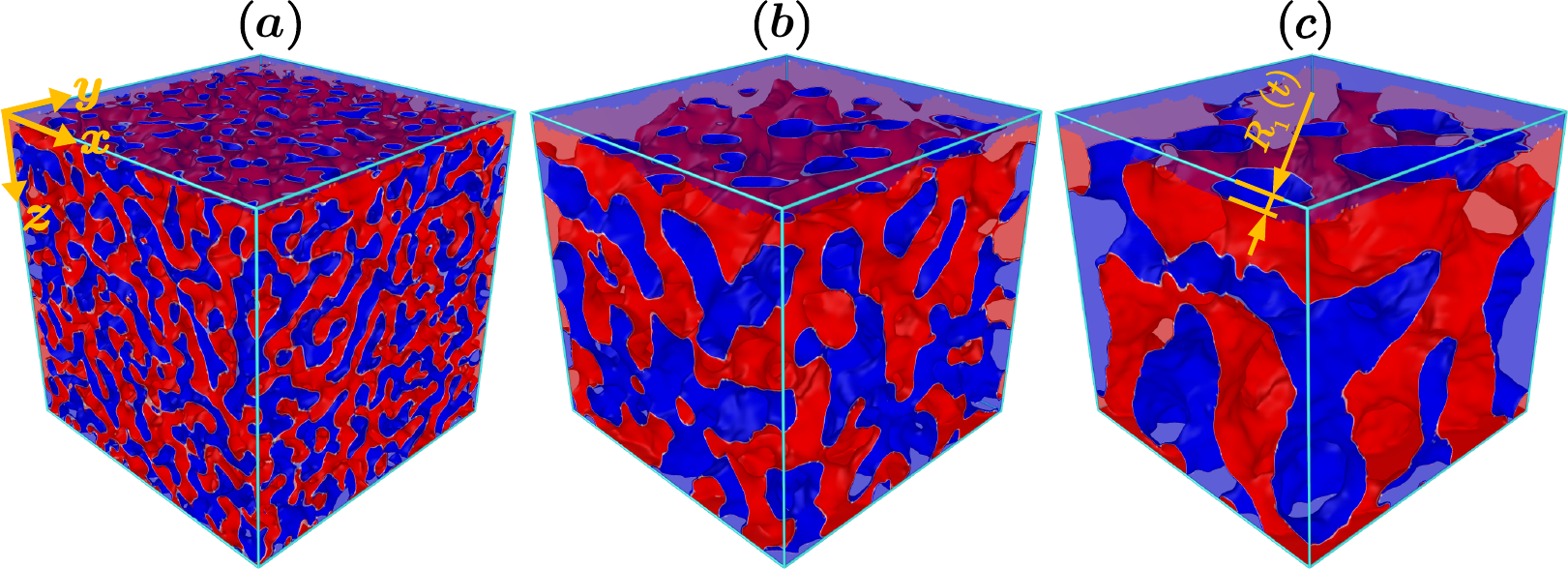}
		\caption{ Time evolution of the domain interfaces resulting from the interplay of wetting and phase separation kinetics in a binary mixture. The two phases in the binary mixture ($A+B$) undergoing SDSD are shown in blue and red transparent voxels without edges. The inner sides of $A$-rich and $B$-rich phase domains are constructed as red ($\psi\to +1$) and blue ($\psi\to -1$) isosurfaces, respectively. The isosurfaces are constructed on the noise-eliminated coarse-grained simulation box composed of the lattice structure (as discussed in the text). We only show the part of the lattice lying above the wetting layer ($z>R_1(t)$) at three different timestamps: $(a)\; 400$, $(b)\; 2000$, and $(c)\;5000$, which range from diffusive growth regime to viscous hydrodynamics regime. The tubular structures are visible across the depletion region with openings toward the wetting layer; with no dominant droplet morphology present above the wetting layer, and the mixture remains purely bicontinuous (interconnected) near the surface and in bulk.}
		\label{fig:figure666b}
	\end{figure}  
    
    The primary reason for the shallow depletion region ($\psi_{\text{av}} (z,t) > -1$) is the formation of an impressive network of tubes of the wetting components ($A$) connecting the bulk domains and the wetting layer. The tubular network in the depletion region is visible in Fig.~\ref{fig:figure666b}, which clearly depicts the tubes with an inner blue surface leading to the wetting layer. We have truncated the isosurfaces depiction just before the wetting layer of thickness $R_1(t)$, leaving the tubes with an opening. The existence of these tubes governs the crossover between the diffusive growth of the wetting layer and the faster growth due to hydrodynamics at the late stages.\cite{ERD11jpcb,GEF94,tanaka01,tanaka93,tanaka93-3,HA97pre} Fig.~\ref{fig:figure666b} also demonstrates that the dominant morphology of the entire system (wetting + depletion + bulk) is bicontinuous (interconnected) for all coarsening regimes ($t>100$). Some droplets near the wetting layer observed in the snapshot of Fig.~\ref{fig:figure666b} (b) are the final stages of the tube necking due to the Rayleigh instability, a hydrodynamic mechanism. These observations led us to invoke the coarsening mechanism given by Scholten for aqueous polymer mixtures, where diffusive coarsening of $\sim t^{1/3}$ is possible for bicontinuous bulk due to hydrodynamics for length scales $L < \Lambda$. For $L > \Lambda$, it is the usual capillary-induced fluid flow from the bulk to the wetting layer through tubes, yielding the faster linear ($\sim t$) growth during the viscous hydrodynamic regime. 

\subsection{The growth of the wetting layer thickness $R_1(t)$}

\begin{figure}[!htb]
		\centering
		\subfloat{\includegraphics[width=0.47\linewidth]{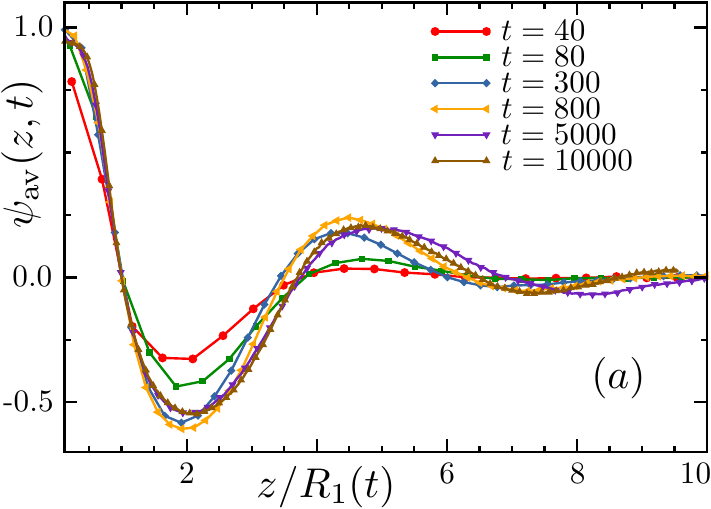}\label{fig:figure2-1a}}\hspace{2mm}
		\subfloat{\includegraphics[width=0.48\linewidth]{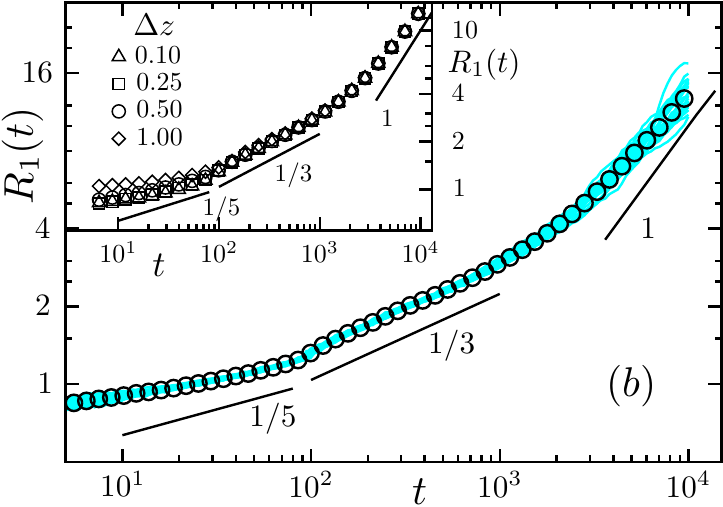}\label{fig:figure2-1b}}
		\caption{($a$) Scaled version of the laterally-averaged order parameter $\psi_{\text{av}(z,t)}$ profiles shown in Fig.~\ref{fig:figure2}. The scaling factor is $R_1(t)$, which is calculated as the first-zero crossing of $\psi_{\text{av}(z,t)}$ along $z$. $(b)$ The wetting layer thickness, $R_1 (t)$, vs. time plotted on a log-log scale. The solid lines have slopes of $1/5$, $1/3$, and $1$, depicting potential-dependent, diffusive, and viscous-hydrodynamics growth. Light colored (cyan) solid lines represent $R_1(t)$ growth computed from different simulation trajectories with an independent initial starting configuration. \emph{Inset:} $R_1(t)$ growth for different slab widths denoting the different levels of precision. }
		\label{fig:figure2-1}
	\end{figure}
The oscillatory composition profiles in Fig.~\ref{fig:figure2} are the SDSD waves.\cite{puri05} These composition waves perform damped oscillations about $\psi_0$ before they vanish while penetrating the bulk, where they encounter randomly oriented composition fluctuations. The characteristic wavelength of such oscillations is derived as $\lambda_c \simeq 2\pi/\xi_B$ using the linear stability analysis of the Cahn-Hilliard equation for SDSD. We verify this by analyzing the scaling behavior of the SDSD profiles in different coarsening regimes. The scaling of SDSD profiles is done using the wetting layer thickness, $R_1(t)$, which is calculated as the \emph{first-zero} crossing of $\psi_{\text{av}}(z,t)$ at each time $t$. In Fig.~\ref{fig:figure2-1}($a$), we plot the scaling behavior of the SDSD profiles, and the profiles show dynamical scaling for two different time regimes for $t>80$. Here, the neat data collapse for the small distances from the wall demonstrates that $R_1(t)$ is the dominant length scale near the wetting wall. However, such scaling is absent for $t<100$ during the very early-time kinetics associated with the first coating layer of the wetting component $A$ on the surface. Interestingly, the dynamic scaling hypothesis applies to cases with isotropic growth in all directions, such as in bulk. However, its extension to systems with a nonvanishing current in the scaling direction (concentration gradients due to broken symmetry) is noteworthy.\cite{GA92} The characteristic wavelength of the SDSD waves as observed from the scaled oscillations in Fig.~\ref{fig:figure2-1a} is about $\simeq 5$, which is not far from the value of $\lambda_c = 2\pi/\xi_B \sim 6.28$ (where $\xi_B$ is the bulk correlation length). The discrepancy may be due to the inappropriate use of bulk values with the surface effects.

In Fig.~\ref{fig:figure2-1}($b$), we plot the wetting layer thickness $R_1(t)$ versus time on a log-log scale, where the slope of this curve yields the power-law exponent for any power-law dependence. Indeed, $R_1(t)$ is observed to follow a power-law growth $\sim t^{\alpha}$ with three growth exponents (slopes) corresponding to different coarsening regimes, where each regime is almost a decade long in time. We plotted our data against the known exponents rather than interpreting them using a log-log scale. We compare these regimes with the solid straight lines of slopes $1/5$ (for $n=3$ in long-range surface potential $V\sim 1/z^{n}$) and $1/3$ marking the potential-dependent and diffusive growth of the PB model,\cite{SKH97,puri05,SSPSP24jcp,ZJPP22pre}, and the line with slope $1$ emphasizing the viscous hydrodynamic regime.\cite{tanaka00jpcm,puri05,RLE91} $R_1(t)$ seems to deviate from the potential-driven growth for early times, and it remains so for about $t<10^2$. The intermediate time sees $R_1(t)$ growth driven by some diffusion process with a growth exponent $\alpha = 1/3$. The late-time coarsening of the wetting layer exhibits an accelerated growth due to capillary effects in the viscous hydrodynamic regime with $\alpha = 1$. In Fig.~\ref{fig:figure2-1b}, we also include, shown in a light color, the wetting-layer thickness computed from each independent trajectory and compared to the ensemble-averaged $R_1(t)$. Independent trajectories and their deviations from the ensemble-averaged $R_1(t)$ reveal the different critical pathways that the coarsening wetting layer could adopt for different growth regimes and how tightly they are bound to domain morphologies. For the diffusive growth stage, the independent curves remain closer to the average one and therefore signify no variation in the morphological features of the system across different simulation runs. However, some deviation is observed in late-time hydrodynamical growth, which depends on the aspect ratio of the tubes connecting the wetting layer and the bulk, as shown in Fig.~\ref{fig:figure666b}. Although these deviations are present for the late time, they all exhibit a growth exponent larger than the diffusive $1/3$. This is clear from the crossovers between the diffusion-dominated and viscous hydrodynamic coarsening regimes shown by the ensemble-averaged $R_1(t)$ and all cyan-colored data.

\subsection{The crossover $\alpha = 1/3 \to 1$ in the wetting-layer growth}
In Fig.~\ref{fig:figure2-1b}, the crossover in $R_1(t)$ from diffusion to the viscous hydrodynamic regime is broad and gradual, pointing to interplay between some competing dynamics. These slow crossovers are also observed in various experimental and numerical investigations for bulk fluid mixtures, as discussed in the Introduction.\cite{ERD11jpcb,HA97pre,ERD12} We attempt to quantify the crossover between the diffusive and viscous growth regimes of $R_1(t)$, as follows. In the intermediate (diffusive) regime, $R_1(t)$ follows
\begin{equation}\label{eq:diff}
R_1(t) \simeq A\,t^{1/3},
\end{equation}
whereas in the viscous (capillary-driven) regime, it approaches a linear form with an offset,
\begin{equation}\label{eq:visc_offset}
R_1(t) -R'_1 \simeq B\,t^{\alpha}, \qquad \alpha\approx 1.
\end{equation}
The crossover time $t_c$ is the point of intersection of the diffusive and linear curves. To robustly determine the offset $R_1'$, we introduce the instantaneous (running) growth exponent
\begin{equation}\label{eq:inst}
\alpha_i(t)\equiv\frac{d\ln R_1(t)}{d\ln t}.
\end{equation}
Differentiating Eq.~\eqref{eq:visc_offset} gives the exact relation
\begin{equation}\label{eq:alphai}
\alpha_i(t)=\alpha\!\left(1-\frac{R_1'}{R_1(t)}\right).
\end{equation}

\begin{figure}[!htbp]
		\centering
		\subfloat{\includegraphics[width=0.4\linewidth]{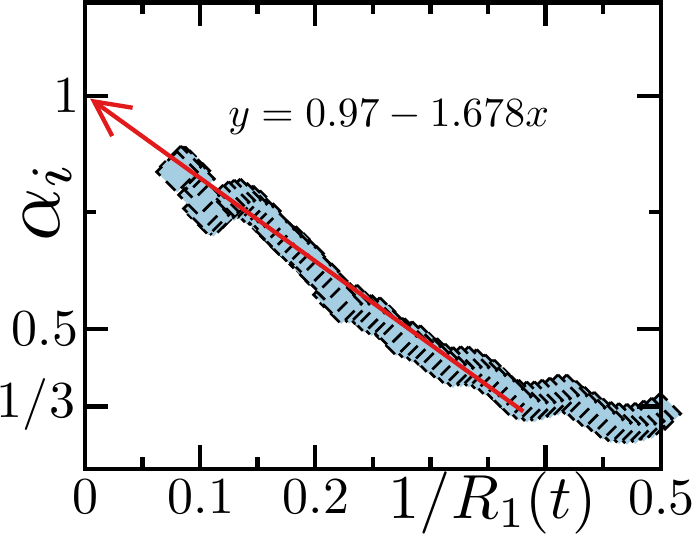}\label{fig:figure5a}}\hspace{5mm}
		\subfloat{\includegraphics[width=0.38\linewidth]{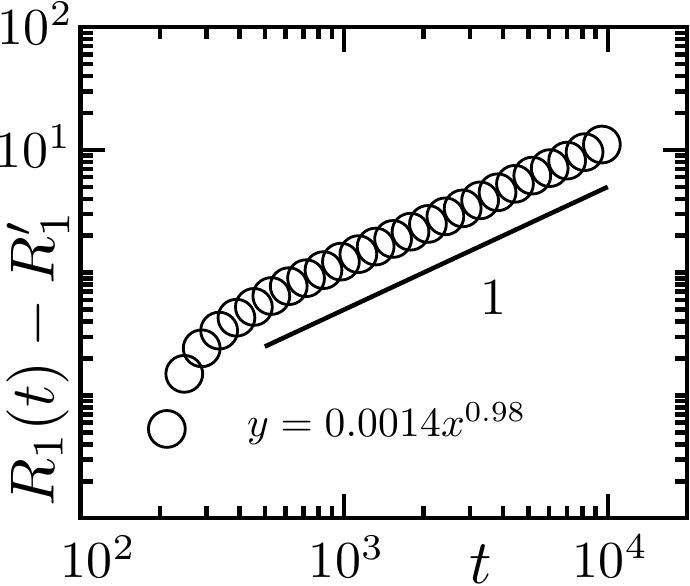}\label{fig:figure5b}}
		\caption{\emph{Left Panel:} Computation of the length offset $R'_1$ from the fitting of the instantaneous exponent $\alpha_i$ as a function of increasing wetting layer thickness $R_1(t)$. \emph{Right Panel:}~Fitting~of~$R_1(t) - R'_1$ vs. simulation time $t$, which exhibits a linear growth.~From fitting of the diffusive coarsening of the wetting layer in Fig.~\ref{fig:figure2-1b}: $y=0.3t^{0.33}$ and the linear growth $y=0.0014x^{0.98}$, and equating $At^{1/3}=Bt$, we get the crossover time of $\approx 3000$. The crossover time is in the range of crossover seen for the bulk domain coarsening (Fig.~\ref{fig:figure1}).\cite{SSS10,SSS12} }
		\label{fig:figure5}
	\end{figure}
Hence $\alpha_i$ depends linearly on $1/R_1(t)$, and a linear fit of $\alpha_i$ versus $1/R_1(t)$ yields the intercept $\alpha$ at $1/R_1\to0$ and the slope $-\alpha R_1'$. From the linear fit in Fig.~\ref{fig:figure5a}, we found the intercept $\alpha\to 1$ as $1/R_1(t)\to 0$ and obtained the offset $R'_1 \simeq 1.7$. We subtract this offset from the measured $R_1(t)$ and fitted the viscous regime to $R_1(t)-R_1'=B\,t^{\alpha}$. After offset removal, the viscous regime is well described by $R_1(t)-R_1'\simeq B\,t^{0.98}$, i.e. consistent with the linear hydrodynamic scaling $\alpha=1$ within uncertainty. Finally, we determine the crossover time $t_c$ by equating the diffusive and viscous forms. Using the fitted values of $A,B$ we obtain $t_c\simeq 3000$ and the corresponding crossover length $R_1(t_c)\approx 5.3$. The crossover time between diffusive and viscous hydrodynamic growth, $t_c \sim 3000$, is similar to the value obtained for bulk studies investigating SD.\cite{SSS10,SSS12}

Numerical investigations exploring bulk domain coarsening and wetting-layer growth in binary fluid mixtures have reported correct growth exponents with crossover ($t^{1/3}\to t$), but have not paid much attention to the underlying physical mechanism. We tend to characterize the wetting-layer growth observed in our simulations by invoking the mechanism proposed by Scholten\cite{ELE05macro38} for the crossover observed in the bulk length scale in polymer mixtures.\cite{NAA01macro34} For completeness, we state the hydrodynamic starting point and the curvature contribution in the interfacial surface tension used in Scholten’s argument. The fluid motion obeys the incompressible Navier–Stokes equation,
\begin{equation}\label{eq:NS}
    \rho\left(\frac{\partial \mathbf v}{\partial t} + \mathbf v\!\cdot\!\nabla\mathbf v\right)
= -\nabla P + \eta\nabla^2\mathbf v,
\qquad \nabla\!\cdot\!\mathbf v=0,
\end{equation}
and in the low–Reynolds number flow, inertia is negligible so that $-\nabla p +\eta\nabla^2\mathbf v=0$ (Stokes flow). The capillary pressure ($P$) driving the coarsening depends on the size of the domains, $L(t)$, and scales as $\Delta P\!\sim\!\gamma(L)/L$. So, the characteristic velocity $(v)$ scale follows from the viscous balance $\eta v/L^2\sim\Delta P/L$, giving $v\sim\gamma(L)/\eta$ and hence
\begin{equation}\label{eq:viscous}
    \frac{dL}{dt}\sim\frac{\gamma(L)}{\eta}.
\end{equation}

The Helfrich bending energy gives the curvature dependence of the interfacial tension
\begin{equation}\label{eq:Helfrich}
    \gamma(J,K)=\gamma_0 + kC_0J + \frac{1}{2}kJ^2 + \Bar{k}K
\end{equation}
in which $J$ and $K$ are the total curvature and the Gaussian curvature. Scholten approximated Eq.~\ref{eq:Helfrich} by assuming spontaneous curvature $C_0$ and rigidity constant $\Bar{k}$ associated with the Gaussian curvature to be zero and neglecting $\Bar{k}K$. Substituting the total curvature to be domain-size dependent, $J\propto L^{-1}$, he came up with the final expression for the interfacial tension
\begin{equation}\label{eq:Helf_approx}
    \gamma(L) = \gamma_0 + \frac{2k}{L^2},
\end{equation}
where $\gamma_0$ is the surface tension for flat interfaces and $k$ is the bending rigidity. We identify these assumptions and the final surface-tension expression with $\gamma(\lambda)$ defined for cylindrical droplets with zero Gussian curvature. 

Substituting Eq.~\ref{eq:Helf_approx} into the hydrodynamic rate equation Eq.~\ref{eq:viscous} yields the two limits quoted by Scholten: for $L\gg \Lambda$ the flat-surface term dominates and $dL/dt\sim\gamma_0/\eta$ gives $L~\sim(\gamma_0/\eta)t$, while for $L\ll \Lambda$ the bending term dominates and $dL/dt\sim 2k/(\eta L^{2})$ results into $L\sim~(6k/\eta)^{1/3}t^{1/3}$. Equating the two contributions defines the crossover length
\begin{equation}\label{eq:crossover}
    \Lambda \;=\; \sqrt{\frac{2k}{\gamma_0}}
\end{equation}
to our interest, differentiating the growth regimes dominated by the bending rigidity ($L \ll \Lambda$), and the stretching ($L \gg \Lambda$) of the domain interfaces. We compute $\Lambda$ for our simulations using $\gamma(L)$ expression given for a symmetric LJ model,\cite{SK11pre84,BSMPK10jcp133,ODD15jcp142} 
\begin{equation}\label{eq:das_gamma}
    \gamma(R)=\frac{\gamma_0}{1+2\frac{\ell^2}{L}},
\end{equation}
where $\ell$ is related to the bending rigidity ($k$) and the rigidity constant associated with the Gaussian curvature ($\Bar{k}$) using the mean field results as\cite{ED93MP80} 
\begin{equation}\label{eq:ell_k}
    \ell^2 = \frac{k+\Bar{k}/2}{\gamma_0}.
    \end{equation}
Das \emph{et. al.}\cite{SK11pre84,BSMPK10jcp133} have extensively studied the dynamical behavior of length $\ell$ for the same symmetric mixtures used in the current study. They found that the length $\ell$ has a critical divergence similar to the bulk correlation length $\xi_B$, but show that its amplitude is significantly larger ($\ell \approx 4\xi_B$).\cite{SK11pre84} For no Gaussian curvature ($\Bar{k}=0$) for cylindrical tubes connecting the bulk and the wetting layer in our simulations, as clearly visible in Fig.~\ref{fig:figure666b}, we obtain the relation 
\begin{equation}\label{eq:ell_xi_k}
    \ell^2=(4\xi_B)^2=k/\gamma_0,
\end{equation}
where $\xi_B \simeq \sigma = 1 $ is the bulk correlation length for the current LJ binary mixture system, which phase separates into two coexisting phases, having the Ising model-like up-down symmetry. Substituting $\xi_B$ in Eq.~\ref{eq:ell_xi_k} we get the ratio $k/\gamma_0 \approx 16$, which we replace in Eq.~\ref{eq:crossover} to get the crossover length $\Lambda \approx 5.7$. This value agrees with the crossover length $ R_c\approx 5.3$ corresponding to the crossover time $t_c\approx 3000$ calculated earlier for $R_1(t)$ in Figs.~\ref{fig:figure2-1b} and~\ref{fig:figure5}. The matching crossover length scales and bicontinuous domain morphology thus verifies the bending-dominated diffusive wetting-layer growth for intermediate times ($t<1000$), which later slowly crosses over to the usual linear viscous hydrodynamic regime. In the linear hydrodynamic regime, the growth follows from the stretching of the interfaces, yielding capillary-induced fluid flow and accelerated coarsening.

\subsection{Fluid Density Distribution in Amorphous versus Flat walls}
	\begin{figure}[!htbp]
		\centering
		\includegraphics[width=0.5\linewidth]{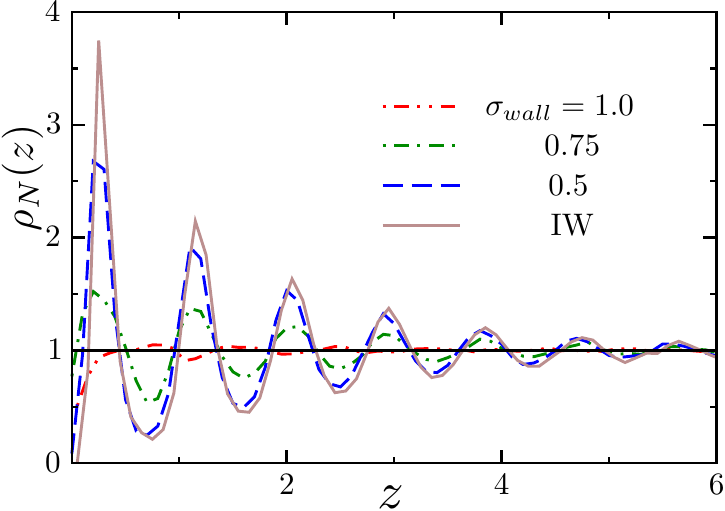}
		\caption{Reduced number density $\rho_N(z)$ plotted as a function of the depth $z$ from the wetting wall. The profiles correspond to different walls made up of particles of size $\sigma_{wall}$. We chose a high packing fraction for $\sigma_{wall}=0.75$ and $0.5$ to reproduce the flat wall conditions. We have also included the density data from the integrated wall for comparison. The profile for $\sigma_{wall}=1$ does not exhibit the non-uniform density oscillations near the wall, eliminating the wall-induced crystallization in the fluid mixture.  }
		\label{fig:figure3}
	\end{figure} 
    Here we briefly investigate how the wall's roughness affects the fluid density next to the walls. In our previous work on SDSD,\cite{ZJPP22pre,PSS12,PSS12-epl} including recent ones,\cite{ZJPP22pre,SSPSP24jcp}, we took the integrated LJ potential (IW) potential to model the wall-fluid interaction, which is defined as:
	\begin{equation}\label{eq1:wall}
	u_w(z)=\frac{2\pi \rho_N \sigma^3}{3}\left[ \frac{2\epsilon_r}{15}\left( \frac{\sigma}{z^\prime}\right)^9 - \delta_\alpha \epsilon_a\left(\dfrac{\sigma}{z^\prime}\right)^3 \right],
	\end{equation}
	where $\alpha \in \{A,B\}$, $\sigma$ is the fluid-wall overlap distance, and $z\prime = z+\sigma$ so that singularity in $u_{w}(z)$ does not fall on the surface at $z=0$. The attractive part of $u_w(z)$ is similar to $V(z)~\sim z^{-n}$ of PB model with $n=3$. The preference of $A$-type particles by the wall materializes by the setting $\delta_A =1$ and $\delta_B=0$. These unstructured walls (flat walls) cause stacking of fluid particles into layers in its vicinity,\cite{PWA93,PWA94prl,SSJ06-pre} which subsequently make it difficult to measure $R_1(t)$ with microscopic precision. To produce similar effects in the amorphous walls, we recreate flat wall conditions by increasing the packing fraction of wall particles. To this end, we simulate smaller systems of size $(32\sigma)^3$ differing in the size of the wall atoms ($\sigma_{wall}$) and with highly packed walls. The resulting number densities appear in Fig.~\ref{fig:figure3}, where we have also shown the density profiles from the IW setup as a reference, which is a densely packed wall of infinite packing fraction.\cite{ZJPP22pre} Clearly, walls with higher volume fractions of particles exhibit number densities closer to the integrated wall. Therefore, by keeping wall density similar to fluid density for $\sigma_{wall}=1$, we avoided the wall-induced particle layering and could access the late-time growth regimes. 

\subsection{Coarsening of Domains Parallel and Perpendicular to the Wall}
We conventionally characterize SDSD using spatial correlation functions to understand the dynamical behavior of characteristic fluctuation (domain coarsening) in cross sections parallel and perpendicular to the wall. To investigate domain coarsening in cross sections parallel to the wall, we define a depth-dependent two-point equal-time layerwise correlation function\cite{puri05,SK94jsp} 
\begin{equation}
\begin{split}
C_{||}(\vec{\rho},z,t) = \langle \psi (\vec{R},z,t)\psi (\vec{R}+\vec{\rho},z,t)\rangle	- \langle \psi (\vec{R},z,t)\rangle \langle \psi(\vec{R}+\vec{\rho},z,t)\rangle,
\label{eq:Clay}
\end{split}
\end{equation}
where, the angular brackets denote the averaging over independent runs and integration over the position vector $\vec{R}$ in layers parallel to the wall. A scalar correlation function $C_{||}(\rho,z,t)$ is obtained after we spherically average the contributions for all $\vec{\rho}$ defined in the $xy$ plane. Switching to cross sections perpendicular to the wall is done by computing correlation functions in the $xz$ planes ($yz$ planes are equally fine due to semi-infinite geometry in the $xy$ direction), which gives $C_{\perp}(\vec{\rho_z},x,t)$. Subsequently, we extract $L_{||}(z,t)$ and $L_{\perp}(x,t)$ from the respective spherically averaged correlation functions as a distance over which the correlation values decay to zero.\cite{puri05,SK94jsp} Furthermore, the correlation functions are calculated from the noise-eliminated coarse-grained simulation box with a lattice structure, whose construction we have discussed earlier, where each lattice point denotes the order parameter $\psi=n_A-n_B$ (see a similar representation in Fig.~\ref{fig:figure666b} where each lattice point is a voxel).

	\begin{figure}[!htbp]
		\centering
		\includegraphics[width=0.6\linewidth]{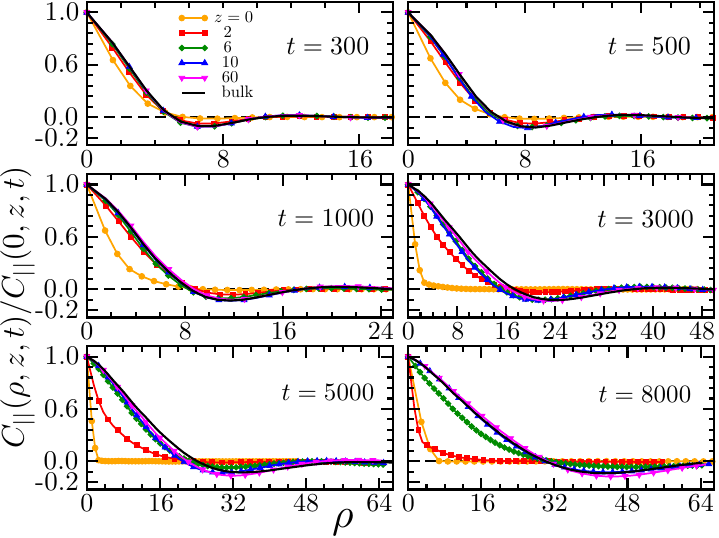}
		\caption{Plots of depth-dependent real-space correlation functions, $C_{||}(\rho,z,t)/C_{||}(0,z,t)$, in the direction parallel to the wetting substrate. Each subplot corresponds to different layers placed at a depth $z$ from the wetting surface. Different subplots contain data sets of $C_{||}(\rho,z,t)/C_{||}(0,z,t)$ at different reduced times specified.}
		\label{fig:figure6unsc}
	\end{figure}

We examine $C_{||}(z,t)$ at multiple depths along the composition wave and compare them with the correlation values from the bulk (isotropic systems without walls). The results are shown in Fig.~\ref{fig:figure6unsc}, where we have carefully chosen layers from the wetting layer, the depletion region, and the bulk. The correlation data corresponding to the bulk layers oscillate about zero, indicating an interconnected domain morphology for a critical composition ($\psi_0=0$). However, $C_{||}(z,t)$ in the wetting layers ($z=0,2$) shows overdamped oscillations resulting from the off-critical domain morphologies, similar to spinodal decompositions in magnetic systems with a non-conserved order-parameter. The pinning of the correlation value in the layers at $z=0$ for time $t>1500$ corresponds to the saturation of $\psi_{\text{av}(z,t)}\simeq 1$ in the wetting layers ($z<R_1(t)$). The pinning is also present in the form of shoulder formation in $\psi_{\text{av}}(z,t)$ as seen in Fig.~\ref{fig:figure2} for profiles close to the wall. Moreover, the underdamped correlations in layers lying above the wetting layers denote the presence of off-critical morphologies from the cross-section of tubes of the wetting phase suspended in the sea of the nonwetting phase. In particular, there are disparities in the decay distances of $C_{||}(z,t)$, suggesting substantial morphological changes in the regions affected by the SDSD waves with large amplitude. The disparities also occur when compared with the bulk correlations; however, they are associated only with the shifting morphologies in the affected cross sections.

 \begin{figure}[!htbp]
		\centering
		\subfloat{\includegraphics[width=0.48\linewidth]{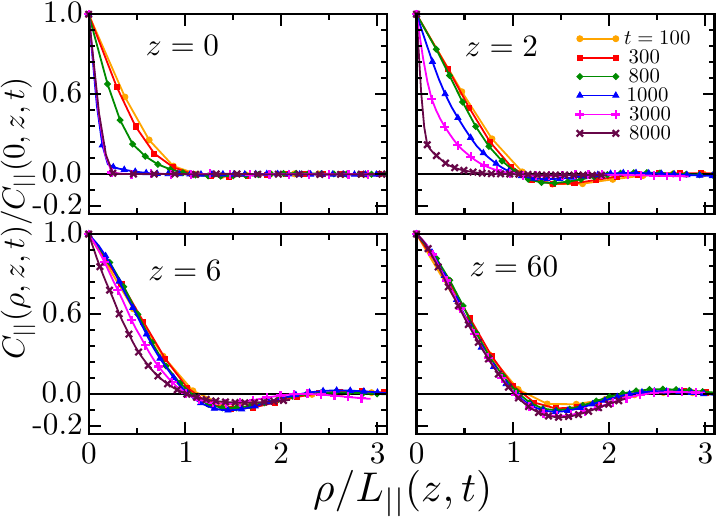}\label{fig:figure6a}}\hspace{2mm}
		\subfloat{\includegraphics[width=0.48\linewidth]{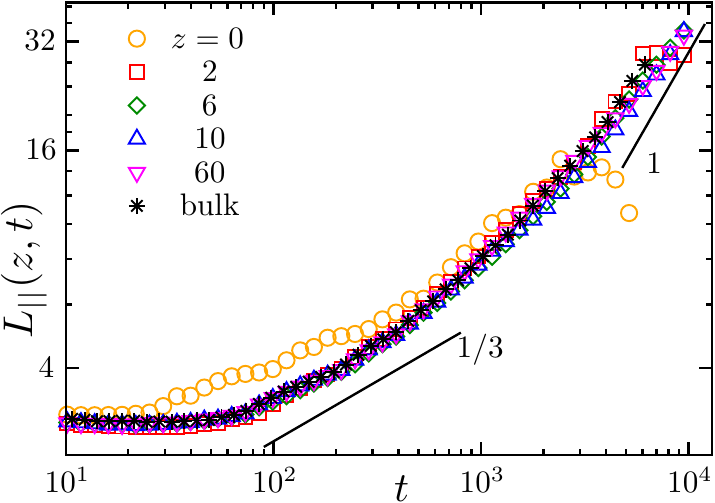}\label{fig:figure6b}}
		\caption{(a) Plots of $C_{||}(\rho,z,t)/C_{||}(0,z,t)$ vs. scaled distances for layers in the vicinity of the wetting surface and from the bulk. Scaling in each subplot for different times is performed using characteristic length scales, $L_{||}(z,t)$, extracted as the decay distance of $C_{||}(\rho,z,t)/C_{||}(0,z,t)$ to $0$. (b) Time dependence of the parallel length scale $L_{||}(z,t)$ at different depths from the wetting surface. The data correspond to growths in layers placed at $z=0, 2,6,10$ (near the wetting surface) and $z=60$ (away from the wetting surface). The bulk data corresponds to the length scale growth in a pure phase-separating system without walls. Solid lines shown here have slopes of $1/3$ and $1$, demonstrating diffusive and viscous hydrodynamic growth regimes. }
		\label{fig:figure6}
	\end{figure}   

In Fig.~\ref{fig:figure6a}, we focus on the scaling behavior of the correlation data as a function of depth $z$ from the surface. The radial distance, $\rho$, is rescaled by the length scale $L_{||}(z,t)$, where the latter is the distance at which $C_{||}(\rho,z,t)$ decays to zero. $C_{||}(\rho,z,t)$ exhibits good data collapse at all distances for two different growth regimes for the bulk layers. Furthermore, no reasonable scaling is seen for the layers lying inside the wetting layer ($z < R_1(t)$) at that time.

	We plotted the extracted $L_{||}(z,t)$ vs. $t$ on a log-log plot in Fig.~\ref{fig:figure6b}. Not surprisingly, the bulk layers show a length scale that grows as $t^{1/3}$ and $t^{1}$, corresponding to diffusion\cite{IV61} and viscous hydrodynamics.\cite{tanaka01,tanaka93} Also, the layers affected by the SDSD waves show similar growth regimes, except for the layers with highly off-critical compositions ($z=0$). Noticeably, the accelerated growth regime by Puri \emph{et al.} for long-range surface fields like ours is absent here. They provided two possible explanations for the accelerated growth, $L_{||}(z,t)\sim t^{1/2}$. Firstly, they considered it a transient regime, where the bicontinuous domain morphology breaks down into isolated elongated droplets of the wetting component, oriented parallel to the wetting layer on average. Such elongated droplets have a stronger tendency to compactify, forming larger domains, resulting in accelerated growth, similar to droplet growth via the Brownian-Coagulation process for two-dimensional systems.\cite{siggia79,ohta84ap,KD74,binder77prb} Furthermore, the layerwise effects were prominent and long-lived due to their greater proximity to the wetting surface and are perpetually propagated into the bulk with penetration of it by SDSD waves. Secondly, a length scale associated with the fluctuations about a highly off-critical background ($\psi_0 \ne 0$) also results in $\alpha = 1/2$. Furthermore, their parallel length scale also showed depth-dependent orientational effects of the wetting layer, which was indicated by the increased growth magnitudes in $L_{||}(z,t)$ for $z$ closer to the wetting wall. Similar elongation of the domains parallel to the wall was qualitatively reported by Brown and Chakrabarti\cite{GA92} in their Monte Carlo simulations of SDSD. However, our MD results do not observe any asymmetry in the dynamical behavior of domain coarsening in the parallel and perpendicular directions above the wetting layer. This is partly confirmed by the overlap of $L_{||}(z,t)$ and the bulk length scale $L(t)$ in Fig.~\ref{fig:figure6b}, which means that the domains just above the depletion layers are isotropic, similar to the bulk domains. 
 \begin{figure}[!htbp]
		\centering
		\includegraphics[width=0.5\linewidth]{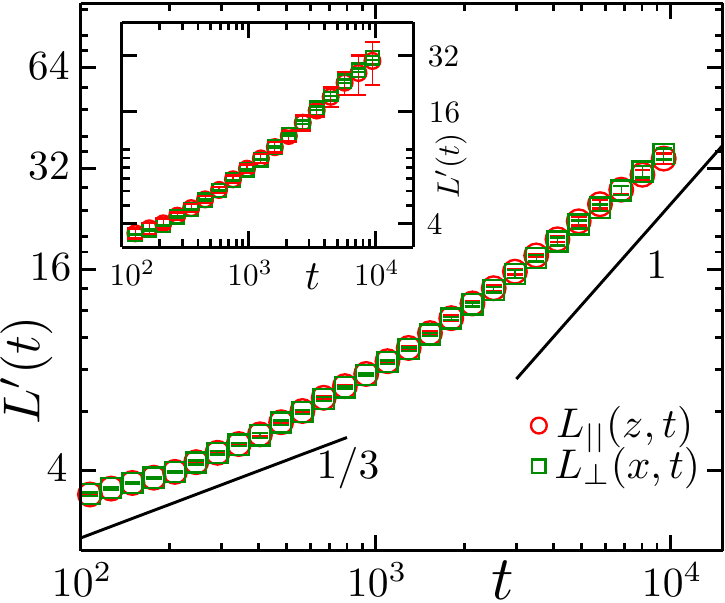}
		\caption{Time evolution of the length scale $L^{\prime}$ computed for cross sections parallel or perpendicular to the wall. The cross sections used in the main panel are of size $64 \times 64$ and start above the wetting layer for the complete simulation time. In the \emph{inset}, we chose bigger cross sections of size $128\times 128$ covering the wetting layer. Solid lines are guidelines with slopes $1/3$ and $1$ representing diffusive and viscous hydrodynamic regimes.    } 
		\label{fig:L2}
    \end{figure}
   
    For better statistics, we investigate the time evolution of $L^\prime (t)$ (that is, $L_{||}$ and $L_{\perp}$) from each independent trajectory and compute their average and standard deviation. The results for both length scales are plotted in Fig.~\ref{fig:L2} for square cross sections of size $64\times 64$ starting at a depth ($z=16$) that is above the wetting layer, $z>R_1(t)$, for the complete simulation time. The figure shows that both length scales parallel and perpendicular to the wall overlap with negligible standard deviation, signifying isotropic domains just above the depletion layers, in contrast to the asymmetric domain coarsening previously reported. However, for cross sections of size $128\times 128$, which also covers the wetting layer, $L_{||}$ in the inset of the figure shows that the standard deviations diverge over time. The divergence was inevitable as some cross sections entered the wetting and depletion regions with a highly off-critical domain morphology. However, they still overlap reasonably well for our simulation time, indicating no significant disparities between the dynamical behavior of domain coarsening in different directions. In addition, these length scales with $R_1(t)$ also do not show \emph{fast-mode kinetics} as reported for the combined effects of hydrodynamics and long-range interactions in low-weight polymer mixtures.     

\subsection{Conclusions}
This paper investigates SDSD in binary fluid mixtures on an amorphous substrate using molecular dynamics simulations. Our results show that the wetting-layer thickness grows as a power law, $R_1(t)\sim t^\alpha$, with $\alpha\approx1/3$ in an intermediate regime that crosses over to a late-stage viscous-hydrodynamic regime with $\alpha\approx1$. The two growth regimes are observed for at least an order of magnitude in time and are separated by a crossover that agrees with previous numerical and experimental studies based on Flory–Huggins theory, polymers/biopolymers, and colloid-polymer mixtures. The roughness of the amorphous wall makes our setup stand out among previous attempts to study SDSD via molecular simulations. The roughness at the fluid particle scale allows us to avoid the layering effects observed in walls modeled using an integrated wall potential (flat walls). For flat walls, the stacking of particles in the layers has caused the fluid mixtures to crystallize once the wetting layer becomes $2\sigma$ to $3\sigma$ thick (where $\sigma$ is the fluid particle diameter). With rougher walls, we could access growths of the wetting layer and bulk domains during the late stages of SDSD, which had not been previously achieved.

We computed the crossover time and the corresponding crossover length for $t^{1/3} \to t$ and found that the crossover of the wetting-layer thickness coincides quantitatively with the crossover that we measure in the bulk mixtures. This coincidence differs from previous observations in some colloid polymer experiments where the wetting layer crossover was reported to occur significantly later than the bulk crossover. The crossover length extracted from our data agrees with the expression derived by Scholten, $R_{c} \approx \Lambda =\sqrt{2k/\gamma_{0}}$ for the bulk crossovers seen in biopolymer mixtures, which follows from a curvature-dependent interfacial tension between the two coexisting phases that includes a bending contribution. This agreement supports the physical picture that, in bicontinuous morphologies and in tube-like connections between bulk domains and the wetting layer, the bending contribution governs small-scale coarsening and gives rise to the $t^{1/3}$ growth, while at larger length scales, capillary-driven viscous flow controls the linear growth regime. Thus, beyond presenting the growth laws, we have addressed the underlying physical mechanism of the diffusive regime in an interconnected mixture where droplet-based evaporation condensation or Brownian coagulation mechanisms are not applicable.

We also examined the coarsening dynamics parallel and perpendicular to the wall and found no significant anisotropy in the growth rates, a result we attribute to the bicontinuous connectivity of the domain morphology, where bending rigidity dominates over the orientational effects of the wall. This absence of an orientational effect contrasts with some continuum model results but is consistent with the domain geometry present in our simulations. In general, our results link microscopic interfacial mechanics and macroscopic coarsening kinetics and provide a coherent explanation for the observed crossover from diffusion to viscous hydrodynamics growth regimes in systems with bicontinuous morphology.

\begin{acknowledgement}
\end{acknowledgement}





\bibliography{main}	

\end{document}